\pdfoutput=1
\RequirePackage{ifpdf}
\ifpdf 
\documentclass[pdftex]{sigma}
\else
\documentclass{sigma}
\fi

\usepackage[mathscr]{eucal}

\numberwithin{theorem}{section}
\newtheorem{cor}[theorem]{Corollary}
\newtheorem{prop}[theorem]{Proposition}
\theoremstyle{definition}
\newtheorem{rem}[theorem]{Remark}
\numberwithin{equation}{section}

\def\R{\mathbb R}
\def\RP{{\mathbb R}\mathbb{P}}

\renewcommand{\L}{\mathcal{L}}
\def\SL{\mathrm{SL}}
\def\M{\mathcal{M}}
\def\Mhat{\widehat M}
\def\P{\mathcal{P}}
\def\calR{\mathcal{R}}
\def\calN{\mathcal{N}}
\def\Rt{\tilde\calR}
\def\Ft{\tilde F}
\def\Q{\mathcal{Q}}
\def\G{\mathcal{G}}
\def\gam{\mbox{\raisebox{.45ex}{$\gamma$}}}
\def\ri{\mathrm{i}}
\def\rd{\mathrm{d}}
\def\dx{\rd{x}}
\def\rhat{\widehat{\rho}}
\def\di{\partial}
\def\transpose#1{{#1}^{\rm T}}

\def\vy{\mathsf y}
\def\E{\mathsf E}
\def\bt{\tilde{b}}
\def\ct{\tilde{c}}
\def\cV{\mathscr{V}}

\begin{document}

\allowdisplaybreaks

\renewcommand{\thefootnote}{$\star$}

\renewcommand{\PaperNumber}{022}

\FirstPageHeading

\ShortArticleName{Integrable Flows for Starlike Curves in Centroaf\/f\/ine Space}

\ArticleName{Integrable Flows for Starlike Curves
\\
in Centroaf\/f\/ine Space\footnote{This paper is a~contribution to the Special Issue ``Symmetries
of Dif\/ferential Equations: Frames, Invariants and~Applications''.
The full collection is available at
\href{http://www.emis.de/journals/SIGMA/SDE2012.html}{http://www.emis.de/journals/SIGMA/SDE2012.html}}}

\Author{Annalisa CALINI~$^{\dag\ddag}$, Thomas IVEY~$^\dag$ and Gloria MAR\'I~BEFFA~$^\S$}

\AuthorNameForHeading{A.~Calini, T.~Ivey and G.~Mar\'i Bef\/fa}

\Address{$^\dag$~College of Charleston, Charleston SC, USA}
\EmailD{\href{mailto:calinia@cofc.edu}{calinia@cofc.edu},
\href{mailto:iveyt@cofc.edu}{iveyt@cofc.edu}}

\Address{$^\ddag$~National Science Foundation, Arlington VA, USA}

\Address{$^\S$~University of Wisconsin, Madison WI, USA}
\EmailD{\href{mailto:maribeff@math.wisc.edu}{maribeff@math.wisc.edu}}

\ArticleDates{Received September 07, 2012, in f\/inal form February 27, 2013; Published online March 06, 2013}

\Abstract{We construct integrable hierarchies of f\/lows for curves in centroaf\/f\/ine $\R^3$
through a~natural pre-symplectic structure on the space of closed unparametrized starlike curves.
We show that the induced evolution equations for the dif\/ferential invariants are closely
connected with the Boussinesq hierarchy, and prove that the restricted hierarchy of f\/lows on
curves that project to conics in $\RP^2$ induces the Kaup--Kuperschmidt hierarchy at the curvature
level.}

\Keywords{integrable curve evolutions; centroaf\/f\/ine geometry; Boussinesq hierarchy;
bi-Hamiltonian systems}

\Classification{37K10; 53A20; 53C44}

\rightline{{\it In honor of Peter Olver.}}

\renewcommand{\thefootnote}{\arabic{footnote}} \setcounter{footnote}{0}

\section{Introduction} \subsection{Integrable evolutions of space curves}

Much of the work on integrable curve evolution equations has been guided by the fundamental role
played by the dif\/ferential invariants of the curve (e.g., curvature and torsion in the Euclidean
setting) in helping identify the curve evolution as an integrable one.
Perhaps the most important example in the case of space curves is that of the Localized Induction
Equation (LIE)
\begin{gather}
\label{LIE}
\gam_t=\gam_{x}\times\gam_{xx},
\end{gather}
describing the evolution of a~curve with position vector $\gam(x,t)$ in $\R^3$, and Euclidean
arclength parameter $x$.
The complete integrability of equation~\eqref{LIE} was uncovered by the realization, due to
Hasimoto~\cite{Ha72}, that the function $\psi=\kappa\exp (\ri\int\tau\,\rd x)$, of
the curvature $\kappa$ and torsion $\tau$ of $\gam$, is a~solution of the cubic focusing nonlinear
Schr\"odinger (NLS) equation
\begin{gather*}
\ri\psi_t+\psi_{xx}+\tfrac{1}{2}|\psi|^2\psi=0,
\end{gather*}
one of the two best-known integrable nonlinear wave equations (the other being the KdV equation).

In this paper, we also use as a~guiding principle the observation that many (but not all)
integrable curve evolutions have the property of {\em local preservation of arclength}, i.e., the
associated vector f\/ields satisfy a~{\em non-stretching} condition.
For example, the LIE vector f\/ield $W=\gam_x\times\gam_{xx}$ satisf\/ies the condition
$\delta_W\|\gam_x\|=0$, where $\delta_W$ denotes the variation in the direction of $W$.
Thus the local arclength parameter $x$ is independent of $t$, and the compatibility conditions
$\gam_{xt}=\gam_{tx}$, $\gam_{xxt}=\gam_{txx}$, $\gam_{xxxt}=\gam_{txxx}$ (more commonly written as
compatibility conditions of the Frenet equations and the evolution equations for the Frenet frame)
turn out to be equivalent to the Lax pair of the NLS equation for $\psi$.

Indeed, many integrable curve evolutions in various geometries have been found by looking for
non-stretching vector f\/ields that produce compatible equations for the moving frame of the
evolving curve; in the case of space curves, the geometries explored include Euclidean~\cite{La77},
spherical~\cite{DoSa}, Minkowski~\cite{N98}, af\/f\/ine and centroaf\/f\/ine~\cite{CQ02}.
(Moreover, integrable curve evolutions {\em without} preservation of arclength have been found in
projective~\cite{M2}, conformal~\cite{M3} and other parabolic geometries.) The approach in these
investigations involves f\/inding suitable choices for the coef\/f\/icients of the non-stretching
vector f\/ields (relative to a~Frenet-type frame) and often assuming special relations among the
dif\/ferential invariants; thus it can be challenging to identify integrable hierarchies.

Another approach to investigating the relation between a~non-stretching curve evolution and the
integrable PDE system satisf\/ied by the dif\/ferential invariants is to seek a~natural Hamiltonian
setting for the curve f\/low.
The LIE was shown by Marsden and Weinstein~\cite{MW} to be a~Hamiltonian f\/low on a~suitable phase
space endowed with a~symplectic form of hydrodynamic origin (see also~\cite{AK, Br}).
In a~fundamental paper~\cite{LP91} Langer and Perline used this framework to explore in depth the
correspondence between the LIE and NLS equations and, along the way, derived a~geometric recursion
operator {\em at the curve level} that made it easy to obtain the integrable hierarchies of both
curve and curvature f\/lows, as well as meaningful reductions thereof~\cite{L99,LP94}.

In this article we study integrable evolution equations for closed curves in centroaf\/f\/ine
$\R^3$ beginning, as in~\cite{LP91}, with a~natural pre-symplectic form on an appropriate
inf\/inite-dimensional phase space.
The Hamiltonian setting allows us to construct integrable hierarchies of curve f\/lows and the
associated families of integrable evolution equations for the centroaf\/f\/ine dif\/ferential
invariants (which turn out to be equivalent to the Boussinesq hierarchies).
The motivation for addressing the centroaf\/f\/ine case comes from an interesting article by
Pinkall~\cite{P95}, who derived a~Hamiltonian evolution equation on the space of closed
nondegenerate curves in the centroaf\/f\/ine plane.
The simple def\/inition of the symplectic form in the planar case (related to the ${\rm SL}(2)$-invariant
area form) suggests that an analogous description may be possible in the $3$-dimensional case,
where a~parallel could be drawn with the more familiar Euclidean case treated by~\cite{LP91}.

Before describing the organization of the paper, we brief\/ly discuss Pinkall's original setting
and some results of ours for the planar case.

\subsection[Pinkall's f\/low in $\R^2$]{Pinkall's f\/low in $\boldsymbol{\R^2}$}

Centroaf\/f\/ine dif\/ferential geometry in $\R^n$ refers to the study of submanifolds and their
proper\-ties that are invariant under the action of $\SL(n)$, not including
translations\footnote{Some authors~\cite{OCent} refer to this geometry as {\it
centro-equi-affine} due to the choice of the unimodular group $\SL(n)$, while using {\it
centro-affine} to refer to geometry invariant under the general linear group $\mathrm{GL}(n)$.}.
For example, a~parametrized curve $\gam:I\to\R^n$ (where $I$ is an interval on the real line) is
{\em nondegenerate} if
\begin{gather*}
\det\big(\gam(x),\gam'(x),\ldots,\gam^{(n-1)}(x)\big)\ne0
\end{gather*}
for all $x\in I$, and this property is clearly invariant under the action of $\SL(n)$.
Thus, for these curves the integral
\begin{gather}
\label{arcint}
\int\big|\gamma,\gamma',\ldots,\gamma^{(n-1)}\big|^{2/n(n-1)}\,\dx
\end{gather}
is $\SL(n)$-invariant, and represents the {\em centroaffine arclength}, where for the sake of
convenience we use the notation
\begin{gather*}
|X_1,\ldots,X_n|:=\det(X_1,\ldots,X_n)
\end{gather*}
for $n$-tuples of vectors $X_i\in\R^n$.
(The fractional power in~\eqref{arcint} is necessary to make the integral invariant under
reparametrization.)

In the case where $n=2$, Pinkall~\cite{P95} def\/ined a~geometrically natural f\/low for
nondegenerate curves in $\R^2$, which he referred to as {\em star-shaped curves}, as follows.
Suppose that $\gam$ is parametrized by centroaf\/f\/ine arclength $s$, so that $|\gam,\gam'|=1$
identically.
It follows that $\gam_{ss}=-p(s)\gam$ where $p(s)$ is def\/ined as the centroaf\/f\/ine curvature.
Along a~closed curve $\gam$, one def\/ines the skew-symmetric form
\begin{gather}
\label{R2pair}
\omega(X,Y)=\oint_\gamma|X,Y|\,\mathrm{d}s,
\end{gather}
where $X$ and $Y$ are vector f\/ields along $\gam$.
This pairing is nondegenerate on the space of vector f\/ields that locally preserve arclength.
Then the symplectic dual with respect to~\eqref{R2pair} of the functional $\oint_\gamma p(s)\,ds$
is the vector f\/ield
\begin{gather*}
X=\tfrac12p_s\gam-p\gam_s.
\end{gather*}

Pinkall's f\/low $\gam_t=\tfrac12p_s\gam-p\gam_s$ induces an evolution equation for curvature that
coincides with the KdV equation, up to rescaling.
In an earlier paper~\cite{CIM}, we showed how to use solutions of the (scalar) Lax pair for KdV to
generate solutions of Pinkall's f\/low.
In particular, we showed that varying the spectral parameter in the Lax pair for a~f\/ixed KdV
potential $q$ corresponds to constructing a~solution to the f\/low with curvature given by
a~Galileian KdV symmetry applied to $q$.
We also derived conditions under which periodic KdV solutions corresponded to smoothly closed loops
(for appropriate values of the spectral parameter) and illustrated this using f\/inite-gap KdV
solutions.

\subsection{Organization of the paper}

 In Section~\ref{S2} we introduce basic notions concerning
the dif\/ferential geometry of nondegenerate curves in centroaf\/f\/ine~$\R^3$, including
centroaf\/f\/ine arclength, dif\/ferential invariants, and non-stretching curve variations.
This section also contains a~discussion of the relation between nondegenerate curves and
parametrized maps into~$\RP^2$.
In Section~\ref{S3} we generalize Pinkall's setting to~$\R^3$ by introducing a~pre-symplectic form
on the space of closed unparametrized starlike curves; we also compute Hamiltonian vector f\/ields
associated with the total length and total curvature functionals.
Flow by these vector f\/ields induces evolution equations for the dif\/ferential invariants; we
discuss these equations in Section~\ref{intsection}, including their bi-Hamiltonian formulation, Lax
representation, and the connection with the Boussinesq equation.
In Section~\ref{Five} we show that the Poisson operators introduced in Section~\ref{intsection} give rise to
the Boussinesq recursion operator, generating a~(double) hierarchy of commuting evolution equations
for the dif\/ferential invariants.
In Theorem~\ref{glorias}, we relate the Hamiltonian structure for starlike curves and the Poisson
structure for the dif\/ferential invariants, and obtain a~double hierarchy of centroaf\/f\/ine
geometric evolution equations.
We conclude Section~\ref{Five}, and the paper, by considering which of these f\/lows preserve the
property that $\gam$ corresponds to a~conic under the usual projectivization map \mbox{$\pi:\R^3\to\RP^2$}.
We show that the sub-hierarchy of conicity-preserving curve evolutions induces the
Kaup--Kuperschmidt hierarchy at the curvature level.

\section[Centroaf\/f\/ine curve f\/lows in $\R^3$]{Centroaf\/f\/ine curve f\/lows in $\boldsymbol{\R^3}$}\label{S2}

\subsection{Centroaf\/f\/ine invariants} Let $\gam:I\to\R^3$ be nondegenerate.
We parameterize $\gam$ by centroaf\/f\/ine arclength, so that
\begin{gather}
\label{unitspeed}
|\gam,\gam',\gam''|=1.
\end{gather}
We assume for the rest of this subsection that $x$ is an arclength parameter.

It follows by dif\/ferentiating~\eqref{unitspeed} with respect to $x$ that
\begin{gather}
\label{gamthird}
\gam'''=p_0\gam+p_1\gam'
\end{gather}
for some functions $p_0(x)$ and $p_1(x)$.
As explained below, these constitute a~complete set of dif\/ferential invariants for nondegenerate
curves.
\begin{rem}
Huang and Singer~\cite{HS} refer to nondegenerate curves in centroaf\/f\/ine $\R^3$ as
{\em starlike}.
They def\/ine invariants $\kappa$ and $\tau$ which correspond to $-p_1$ and $p_0$ respectively.
Labeling $p_0$ as torsion is appropriate, since nondegenerate curves that lie in a~plane in $\R^3$
(not containing the origin) are exactly those for which $p_0$ is identically zero.
\end{rem}
\begin{rem}
\label{projremark}
Some insight into the meaning of the centroaf\/f\/ine curve invariants can be gained by considering
the relationship between $\gamma$ and the corresponding parametrized curve $\Upsilon=\pi\circ\gam$
in $\RP^2$, where $\pi:\R^3\to\RP^2$ is projectivization.
The nondegeneracy condition on $\gam$ corresponds to $\Upsilon$ being regular and free of
inf\/lection points.
Conversely, any such parametrized curve $\Upsilon:\R\to\RP^2$ has a~unique lift to $\gam:\R\to\R^3$
which is centroaf\/f\/ine arclength-parametrized; we refer to $\gam$ as the {\em canonical lift} of
$\Upsilon$.
When written in terms of $\Upsilon$ instead of $\gam$, the invariants $p_0$ and $p_1$ are (up to
sign) the well-known Wilczynski invariants~\cite{Wi}.
Since these invariants def\/ine a~dif\/ferential equation whose solution determines the curve
uniquely up to the action of the group $\SL(3)$, any other dif\/ferential invariant must be
functionally dependent on $p_0$, $p_1$ and their $x$-derivatives.

According to Ovsienko and Tabachnikov~\cite{OT}, the cubic dif\/ferential
$(p_0-\tfrac12p_1')(\dx)^3$ has the interesting property that it is invariant under
reparametrizations of $\Upsilon$.
Curves in $\RP^2$ for which this dif\/ferential vanishes identically are conics.
For curves for which the coef\/f\/icient $p_0-\tfrac12p_1'$ is nowhere vanishing, one can def\/ine
the {\em projective arclength} dif\/ferential $(p_0-\tfrac12p_1')^{1/3}\dx$.
Those parametrized curves in $\RP^2$ for which $p_0-\tfrac12p_1'=C$ (a nonzero constant) are
parametrized proportional to projective arclength, and we use the same terminology for their
canonical lifts into~$\R^3$.
(Note that, in this case, the projective arclength dif\/ferential is~$C^{1/3}$ times the
centroaf\/f\/ine arclength dif\/ferential~$\rd x$.)
\end{rem}

Along a~nondegenerate curve, an analogue of the Frenet frame is provided by vectors
$\gam$,~$\gam'$,~$\gam''$.
In fact, if we combine them as columns in an $\SL(3)$-valued matrix $W=(\gam,\gam',\gam'')$, then
the analogue of the Frenet equations is
\begin{gather*}
W_x=W
\begin{pmatrix}
0&0&p_0
\\
1&0&p_1
\\
0&1&0
\end{pmatrix}
.
\end{gather*}
However, for later use it will be convenient to def\/ine a~dif\/ferent $\SL(3)$-valued frame
$F(x)=(\gam,\gam',\gam''-p_1\gam)$ which satisf\/ies the Frenet-type equation
\begin{gather}
\label{Kfrenet}
F_x=F K,
\qquad
K=
\begin{pmatrix}
0&k_1&k_2
\\
1&0&0
\\
0&1&0
\end{pmatrix},
\end{gather}
where
$k_1=p_1$,
$k_2=p_0-p_1'$.
Of course, $k_1$, $k_2$ also constitute a~complete set of dif\/ferential invariants, and we will come
to use these in place of the Wilczynski invariants from Section~\ref{intsection} onwards.

\subsection{Non-stretching variations}
Suppose that $\Gamma:I\times(-\epsilon,\epsilon)\to\R^3$ is a~smooth mapping such that, for f\/ixed
$t$, $\Gamma(x,t)$ is a~nondegenerate curve parametrized by~$x$.
Without loss of generality, we will assume that $\gam(x)=\Gamma(x,0)$ is parametrized by
centroaf\/f\/ine arclength.
Let $X$ denote the variation of $\gam$ in the $t$-direction, and expand
\begin{gather*}
X=\left.\dfrac{\di}{\di t}\right|_{t=0}
\Gamma=a\gam+b\gam'+c\gam''.
\end{gather*}
(We still use primes to denote derivatives with respect to~$x$, although~$x$ is not necessarily an
arclength parameter along the curves in the family for $t\ne0$.)

To compute the variation of the arclength dif\/ferential $|\gam,\gam',\gam''|^{1/3}\,\dx$, we
introduce the notation $\delta$ for variation in the $t$-direction along $\gam$.
Using the relation~\eqref{gamthird}, we compute
\begin{gather*}
\delta{\gam'}=X'=(a'+p_0c)\gam+(a+b'+p_1c)\gam'+(b+c')\gam'',
\\
\delta{\gam''} =X''=(a''+2p_0c'+p_0'c+p_0b)\gam+(2a'+p_0c+b''+2p_1c'+p_1'c+p_1b)\gam'
\\
\hphantom{\delta{\gam''} =X''=}{}
+(a+2b'+p_1c+c'')\gam''.
\end{gather*}
Then
\begin{gather*}
\delta|\gam,\gam',\gam''|=
|\delta{\gam},\gam',\gam''|+|\gam,\delta{\gam'},\gam''|+|\gam,\gam',\delta{\gam''}|=
3a+3b'+c''+2p_1c.
\end{gather*}
In particular, the variation $X$ preserves the centroaf\/f\/ine arclength dif\/ferential if and
only if
\begin{gather}
\label{nostretch}
b'=-a-\tfrac{1}{3}(c''+2p_1c),
\end{gather}
i.e.,
\begin{gather}
\label{tvector}
X=a\gam-\left(\int\left(a+\tfrac{1}{3}c''+\tfrac{2}{3}p_1c\right)\rd x\right)\gam'+c\gam''.
\end{gather}
We refer to vector f\/ields of this form as {\em non-stretching}, since not only do such variations
preserve the overall arclength of, say, a~closed loop, but also no small portion of the curve is
stretched or compressed.

\section{Hamiltonian curve f\/lows}\label{S3}

\subsection{Symplectic structure on starlike loops}
\label{sympsec}
Generalizing Pinkall's setting~\cite{P95} for planar star-shaped loops to the three-dimensional
case, we introduce the inf\/inite-dimensional space
\begin{gather*}
\widehat{M}=\{\gam:S^1\rightarrow\R^3\,:\,|\gam,\gam',\gam''|=1\},
\end{gather*}
as a~subset of the vector space $V=\operatorname{Map}(S^1,\R^3)$ of $C^\infty$ maps from $S^1$ to
$\mathbb{R}^3$.
Assume that $\gam\in\widehat{M}$, i.e., $\gam$ is a~closed starlike curve parametrized by
centroaf\/f\/ine arclength; then a~vector f\/ield $X=a\gam+b\gam'+c\gam''$ is in the tangent space
$T_\gamma\widehat{M}$ if and only if $X$ is of the form~\eqref{tvector}, where the coef\/f\/icients
$a$ and $c$ are $2\pi$-periodic functions of $x$ and satisfy the ``zero mean" condition
\begin{gather*}
\oint_\gamma\big(a+\tfrac{2}{3}p_1c\big)\,\rd x=0,
\end{gather*}
ensuring that the coef\/f\/icient of $\gam'$ in~\eqref{tvector} is also periodic.

On $V$, def\/ine the skew-symmetric form
\begin{gather}
\label{curvesymplectic}
\omega_\gamma(X,Y)=\oint_\gamma|X,\gam',Y|\,\rd x,
\qquad
X,Y\in T_\gamma V.
\end{gather}
Note that $\omega$ is automatically closed (that is, $\rd\omega=0$) since the integrand
in~\eqref{curvesymplectic} is a~volume form on $\mathbb{R}^3$~\cite{AK,Br}.

Letting $X=a\gam+b\gam'+c\gam''$, $Y=\tilde{a}\gam+\tilde{b}\gam'+\tilde{c}\gam''$, we compute
\begin{gather}
\label{XYcomp}
\omega_\gamma(X,Y)=\oint_\gamma(a\tilde{c}-\tilde{a}c)\,\rd x.
\end{gather}
Assuming that $\gam\in\widehat{M}$ and $X,Y\in T_\gamma\Mhat$, then $\omega_\gamma(X,Y)=0$ for all
$Y$ if and only if $\displaystyle a=-\tfrac{2}{3}p_1c$ and $c$ is constant.
Thus, the restriction of $\omega$ to $\widehat{M}$ is a~degenerate closed 2-form (a pre-symplectic
form), with kernel given by the subspace $\R Z_0+\R Z_1$ of constant-coef\/f\/icient linear
combinations of the vector f\/ields $Z_0=\gam'$ and $\displaystyle Z_1=\gam''-\tfrac{2}{3}p_1\gam$
(corresponding to $c=0$ and $c=1$ respectively).

Note that this degeneracy is the result of restricting the 2-form~\eqref{curvesymplectic} to the space
of closed curves satisfying the arclength constraint.
Degeneracy coming from constraints is common when def\/ining symplectic structures on loop
spaces~\cite{Mo95} and phase spaces of nonlinear evolution equations~\cite{Dorf93}.
\begin{rem} The generators $Z_0$ and $Z_1$ of the kernel of $\omega$ will turn out to be the seeds
of the double hierarchy of curve f\/lows discussed in Section~\ref{Five}.
A similar situation is encountered for the LIE hierarchy~\cite{LP91}, where the seed $\gamma'$ spans
the kernel of the natural pre-symplectic form on loops in Euclidean~$\R^3$.\end{rem}

One could attempt to remove the degeneracy by constructing a~quotient of $\Mhat$ with respect to
group actions generated by f\/lowing by $Z_0$ and $Z_1$.
Flow by $Z_0$ generates an action of the additive group $\R$, simply by translation in $x$.
The resulting quotient space $M=\widehat{M}/\R$ can be identif\/ied with the space of
unparametrized starlike loops.
Because of translation invariance, $\omega$ descends to a~give a~well-def\/ined closed 2-form on
$M$, with one-dimensional kernel $\R Z_1$.
On the other hand, we do not know of a~natural geometric interpretation for the quotient of $M$ by
f\/low under $Z_1$, on which $\omega$ would become non-degenerate.

However, $\omega$ can still be used to def\/ine a~link between vector f\/ields and functionals, and
we will see that $Z_1$ is linked in this way to the arclength functional.

\subsection{Examples}
\label{hamex}
Recall that the correspondence between vector f\/ields $X_H$ and (dif\/ferentials of) Hamiltonians
$H\in C^\infty(M)$ on a~manifold $M$ with symplectic form $\omega$ is def\/ined by the relation
\begin{gather}
\label{hamvf}
\rd H[X]=\omega_\gamma(X,X_H),
\qquad
\forall\, X\in T_\gamma M,
\end{gather}
$X_H$ being the Hamiltonian vector f\/ield corresponding to~$H$.
However, when~$\omega$ is degenerate the correspondence is no longer an isomorphism: for those
functionals~$H$ for which there is a~Hamiltonian vector f\/ield, $X_H$ is only def\/ined up to
addition of elements in the kernel of $\omega$.

We will use~\eqref{hamvf} to compute Hamiltonian vector f\/ields for a~few interesting functionals;
to do so, we will initially work in the ambient space $V$, and use the arclength-preserving
condition~\eqref{nostretch} to rewrite the dif\/ferential in a~form suitable for
applying~\eqref{hamvf}.

We f\/irst consider the arclength functional
\begin{gather}
\label{Ldef}
{L}(\gam)=\oint_{\gamma}\big|\gam,\gam',\gam''\big|^{1/3}\rd x.
\end{gather}
on the space $V$.
Given an arbitrary vector f\/ield $X=a\gam+b\gam'+c\gam''$ (not necessarily arclength preserving),
the variation of the determinant in~\eqref{Ldef} along $X$ is given by
\begin{gather*}
\delta|\gam,\gam',\gam''|=|X,\gam',\gam''|+|\gam,X',\gam''|+|\gam,\gam',X''|=
(3a+3b'+c''+2p_1c)|\gam,\gam',\gam''|.
\end{gather*}
Assume now that $\gam\in\Mhat$, so that $|\gam,\gam',\gam''|=1$ and
\begin{gather*}
\rd{L}[X]=\oint_\gamma\big(a+\tfrac{2}{3}p_1c\big)\,\rd x.
\end{gather*}
We now seek a~vector f\/ield $X_L=\tilde{a}\gam+\tilde{b}\gam'+\tilde{c}\gam''\in T_\gamma\Mhat$
such that $\rd L[X]=\omega_\gamma(X,X_L)=\oint_\gamma(a\tilde{c}-\tilde{a}c)\rd x.
$ Using the non-stretching condition~\eqref{nostretch}, we obtain the following Hamiltonian vector
f\/ield
\begin{gather*}
X_L\equiv Z_1=\gam''-\tfrac{2}{3}p_1\gam
\end{gather*}
(which is unique only up to adding a~constant times $Z_0$).
In Section~\ref{connect} we will see that the associated curve f\/low $\gam_t=X_L$ leads to the
Boussinesq equation for the curvatures $k_1$, $k_2$.

Next, we introduce the total curvature functional
\begin{gather}
\label{funP}
P(\gam)=\oint_\gamma p_1\,\rd x,
\qquad
\gam\in\Mhat.
\end{gather}
From $\gam'''=p_0\gam+p_1\gam'$ and~\eqref{unitspeed}, it follows that $p_1=|\gam,\gam''',\gam''|$.
Then the variation of $p_1$ along an arbitrary vector f\/ield $X=a\gam+b\gam'+c\gam''$ is given by
\begin{gather}
\delta{p}_1= |X,\gam''',\gam''|+|\gam,X''',\gam''|+|\gam,\gam''',X''|
\nonumber\\
\hphantom{\delta{p}_1}{}
= |X,p_0\gam+p_1\gam',\gam''|+|\gam,X''',\gam''|+|\gam,p_0\gam+p_1\gam',X''|
\nonumber\\
\hphantom{\delta{p}_1}{}
= 3p_1a+3c'p_0+2cp_0'+3a''+b'''+4c''p_1+3c'p_1'+cp_1''+5b'p_1+bp_1'+2cp_1^2.\label{deltap1}
\end{gather}
Then, up to perfect derivatives,
\begin{gather*}
\rd P[X]=\oint_\gamma\delta{p}_1\,\rd x=
\oint_\gamma3p_1a+3c'p_0+2cp_0'+4c''p_1+3c'p_1'+cp_1''+4b'p_1+2cp_1^2\,\rd x.
\end{gather*}

Assuming $X$ is an arclength-preserving vector f\/ield, we set $b'=-a-\tfrac{1}{3}(c''+2p_1c)$ and
compute
\begin{gather*}
\oint_\gamma\delta{p}_1\,\rd x=
\oint_\gamma-ap_1+3c'p_0+2cp'_0+\tfrac{8}{3}c''p_1+3c'p_1'+cp_1''-\tfrac{2}{3}cp_1^2\,\rd x.
\end{gather*}
Integrating by parts, we arrive at
\begin{gather}
\label{dP}
\rd P(X)=\oint_\gamma-p_1a+\big({-}p_0'+\tfrac{2}{3}p_1''-\tfrac{2}{3}p_1^2\big)c\,\rd x.
\end{gather}
Suppose that $X_P=\tilde{a}\gam+\tilde{b}\gam'+\tilde{c}\gam''$ is also arclength-preserving.
Setting the right-hand side of~\eqref{dP} equal to
$\displaystyle\omega_\gamma(X,X_P)=\oint_\gamma(a\tilde{c}-\tilde{a}c)\,\rd x$, we get
$\displaystyle\tilde{a}=p_0'-\tfrac{2}{3}p_1''+\tfrac{2}{3}p_1^2$ and $\tilde{c}=-p_1$.
Using equation~\eqref{nostretch} we compute
$\displaystyle\tilde{b}'=-\big(p_0'-\tfrac{2}{3}p_1''+\tfrac{2}{3}p_1^2\big)
-\tfrac{1}{3}\big({-}p_1''-2p_1^2\big)=\big(p_1'-p_0\big)'$,
a~perfect derivative.
Thus, a~Hamiltonian vector f\/ield corresponding to~\eqref{funP} is
\begin{gather*}
X_P=\left(\tfrac{2}{3}\big(p_1^2-p_1''\big)+p_0'\right)\gamma+(p_1'-p_0)\gamma'-p_1\gamma''.
\end{gather*}
(This is only unique up to adding a~linear combination of $Z_0$ and $Z_1$.) Again, we will see that
the associated curve f\/low $\gam_t=X_P$ is also directly related, at the level of the curvatures,
to one of the f\/lows in the Boussinesq hierarchy.

\section{Integrable centroaf\/f\/ine curve f\/lows}
\label{intsection}

In this section we will examine the evolution of centroaf\/f\/ine curvatures induced by the curve
f\/lows def\/ined in Section~\ref{hamex}.
We begin by computing the evolution of invariants under more general curve f\/lows.

\subsection{Evolution of invariants} First, we consider how the centroaf\/f\/ine invariants of
a~starlike curve evolve under a~general non-stretching evolution equation
\begin{gather}
\label{generalflow}
\gam_t=r_0\gam+r_1\gam'+r_2\gam''
\end{gather}
with $r_0=-r_1'-\tfrac13(r_2''+2p_1r_2)$.
(Thus, we assume from now on that $\gam(x,t)$ is parametrized by arclength $x$ at each time.) Of
course, in order for~\eqref{generalflow} to represent a~{\em geometric} evolution equation, $r_1$
and $r_2$ should be functions of the invariants $p_0,p_1$ and their arclength derivatives.
\begin{prop}
The evolution equations induced by~\eqref{generalflow} for the Wilzcynski invariants are
\begin{gather}
(p_0)_t =  -r_1''''+p_1r_1''+3p_0 r_1' +p_0' r_1
\notag\\
\hphantom{(p_0)_t =}{} -\tfrac13\big(r_2'''''+p_1r_2'''\big)+\big((3p_0-2p_1')r_2'\big)'+\tfrac23\big(p_1^2 r_2' +(p_1p_1'-p_1''')r_2\big)+
p_0''r_2,\label{p0genq}
\\
\label{p1genq}
(p_1)_t =  -2r_1'''+2p_1 r_1'+p_1' r_1 -r_2''''+p_1 r_2''+(3p_0 -p_1')r_2' +(2p_0'-p_1'')r_2.
\end{gather}
\end{prop}
\begin{proof}
The second equation~\eqref{p1genq} follows by substituting
$a=-r_1'-\tfrac13(r_2''+2p_1r_2)$, $c=r_2$ in the last line of~\eqref{deltap1}.
Similarly, using $p_0=|\gam''',\gam',\gam''|$, we obtain~\eqref{p0genq} by computing
\begin{gather*}
(p_0)_t=\big|(\gam_t)''',\gam',\gam''\big|+\big|p_0\gam+p_1\gam',(\gam_t)',\gam''\big|+p_0\big|\gam,\gam',(\gam_t)''\big|.\tag*{\qed}
\end{gather*}
\renewcommand{\qed}{}
\end{proof}
We note that these evolution equations previously appeared in~\cite{CQ02}.

From now on, we will take $k_1=p_1$ and $k_2=p_0-p_1'$ as fundamental invariants; one reason for
doing this is that the evolution equations for these invariants induced by~\eqref{generalflow}
take the form
\begin{gather}
\label{kgenflow}
\begin{pmatrix}
k_1
\\
k_2
\end{pmatrix}
_t=\P
\begin{pmatrix}
r_1
\\
r_2
\end{pmatrix}
,
\end{gather}
where $\P$ is the skew-adjoint matrix dif\/ferential operator
\begin{gather}
\label{defofP}
\P=
\begin{pmatrix}
-2D^3+D k_1+k_1D&-D^4+D^2k_1+2D k_2+k_2D
\\[2pt]
D^4-k_1D^2+2k_2D+D k_2&\tfrac23\big(D^5+k_1Dk_1-k_1D^3-D^3k_1\big)+\big[k_2,D^2\big]
\end{pmatrix}
,
\end{gather}
$D$ stands for the derivative with respect to $x$ and $[\cdot,\cdot]$ denotes the commutator on
pairs of operators\footnote{Note that expressions like $Dk_1$ and $Dk_2$ denote composition of $D$ with
multiplication by $k_1$ and $k_2$, respectively.
The skew-adjointness of $\P$ is easy to check, given that $D$ is skew-adjoint.}. This operator $\P$,
which arises naturally when using $k_1$, $k_2$ instead of $p_0$, $p_1$, will play a~signif\/icant role
in the integrable structure of the f\/lows we study.

\subsection{Two integrable f\/lows}
\label{inty}
The vector f\/ield $X_L$ induces a~non-stretching evolution equation
\begin{gather}
\label{B1curv}
\gam_t=\gam''-\tfrac23k_1\gam.
\end{gather}
(This will be the f\/irst non-trivial curve evolution in the hierarchy discussed in
Section~\ref{Five}, where the right-hand side is labeled as $Z_1$.) By setting $r_1=0$, $r_2=1$
in~\eqref{kgenflow}, we obtain the corresponding curvature evolution
\begin{gather}
\label{B1kev}
\begin{pmatrix}
k_1
\\
k_2
\end{pmatrix}
_t=\P
\begin{pmatrix}
0
\\
1
\end{pmatrix}
=
\begin{pmatrix}
k_1''+2k_2'
\\
\tfrac23(k_1k_1'-k_1''')-k_2''
\end{pmatrix}
.
\end{gather}
This PDE system for curvatures is Hamiltonian, since it can be written in the form
\begin{gather*}
\begin{pmatrix}
k_1
\\
k_2
\end{pmatrix}
_t=\P\E k_2,
\end{gather*}
where $\E$ denotes the vector-valued Euler operator
\begin{gather}
\label{Eulerdef}
\E f=
\transpose{\left(\sum_{j\ge0}(-D)^j\dfrac{\di f}{\di k_1^{(j)}},\;\sum_{j\ge0}(-D)^j\dfrac{\di f}{\di k_2^{(j)}}\right)}
\end{gather}
on scalar functions $f$ of $k_1$, $k_2$ and their higher $x$-derivatives $k_1^{(j)}$, $k_2^{(j)}$.
(One can check that the Poisson bracket def\/ined using the Hamiltonian operator~$\P$ on the
appropriate function space~-- see Section~\ref{Psect} below~-- satisf\/ies the usual requirements of
skew-symmetry and the Jacobi identity.)

Moreover,~\eqref{B1kev} can also be written in Hamiltonian form as
\begin{gather}
\label{B1kevQ}
\begin{pmatrix}
k_1
\\
k_2
\end{pmatrix}
_t=\Q\,\E\rho_3,
\end{gather}
for a~dif\/ferent Hamiltonian operator and density
\begin{gather}
\label{defofQ}
\Q=
\begin{pmatrix}
0&D
\\
D&0
\end{pmatrix}
,
\qquad
\rho_3:=\tfrac13(k_1')^2+k_2k_1'+k_2^2+\tfrac19k_1^3.
\end{gather}
(The notation $\rho_3$ is explained below.) Since the curvature evolution can be written in
Hamiltonian form in two ways~\eqref{B1kev} and~\eqref{B1kevQ}, the integrals $\int  k_2\,\dx$ and
$\int \rho_3\,\dx$ are conserved by the f\/low (for appropriate boundary conditions).
\begin{rem} In fact, the curvature evolution here is a~{\em bi-Hamiltonian system}, because~$\P$
and~$\Q$ are a~{\em Hamiltonian pair}, i.e., their linear combinations form a~pencil of Hamiltonian
operators, and a~pencil of compatible Poisson structures.
This assertion can be verif\/ied mechanically (see, e.g., Section~7.1 in~\cite{OLie} for details),
but it also follows from the fact that, at least in the periodic case, the Poisson structures are
reductions of a~well-known compatible pencil of Poisson brackets on the space of loops in
$\mathfrak{sl}(3)$.
(Indeed, when $\gam$ is periodic~--or more generally has monodromy~-- the matrix~$K$ in~\eqref{Kfrenet} provides a~lift into this loop
space.) The proof of the reduction of these brackets can be found in~\cite{DS}, where~$\P$ is linked to the Adler--Gel'fand--Dikii bracket for~$\mathfrak{sl}(3)$ and~$\Q$ is associated to its companion. The brackets were later linked to curve evolutions and dif\/ferential invariants in~\cite{M1}, where more details are available.
\end{rem}

The (negative of the) Hamiltonian vector f\/ield $X_P$ of Section~\ref{hamex} induces the non-stretching evolution
\begin{gather}
\label{B2curv}
\gam_t=k_1\gam''+k_2\gam'+r_0\gam,
\end{gather}
where
\begin{gather}
\label{B2rzero}
r_0=-\big(k_2'+\tfrac13\big(k_1''+2k_1^2\big)\big).
\end{gather}
(The right-hand side of~\eqref{B2curv} is labeled as $Z_2$ in the hierarchy discussed in
Section~\ref{Five}.) We similarly obtain the curvature evolution equations induced by this f\/low
by setting $r_1=k_2$, $r_2=k_1$ in~\eqref{kgenflow}.
We remark that the resulting system is also bi-Hamiltonian, since it can be written as
\begin{gather}
\label{B2kev}
\begin{pmatrix}
k_1
\\
k_2
\end{pmatrix}
_t=\P\,\E\rho_2=\Q\,\E\rho_4
\end{gather}
for
\begin{gather*}
\rho_2=k_1k_2,
\qquad
\rho_4=\tfrac13(k_1'')^2+k_1''\big(k_2'-k_1^2\big)-k_1(k_1')^2+(k_2')^2-k_1^2k_2'+\tfrac19k_1^4+2k_1k_2^2.
\end{gather*}
Thus, $\int \rho_2\,\dx$ and $\int \rho_4\,\dx$ are conserved integrals for~\eqref{B2curv}.
(Because $X_P$ corresponds symplectically to the Hamiltonian $\int  k_1\,\dx$, it is automatic that
this integral is also conserved.)
\begin{rem}
The arclength normalization~\eqref{unitspeed} is preserved by the simultaneous rescaling
$x\mapsto\lambda x$, $\gam\mapsto\lambda^{-1}\gam$.
Under this rescaling, $k_1$ and $k_2$ scale by multiples~$\lambda^2$ and~$\lambda^3$ respectively.
Thus, we may assign {\em scaling weights}~2 and~3 respectively to these curvatures, and each
$x$-derivative taken increases weight by one.

It will turn out (see Section~\ref{curse} below) that the conserved densities for evolution
equations~\eqref{B1curv} and~\eqref{B2curv} are all of homogeneous weight, with one density for
each positive weight not congruent to 1 modulo 3.
We will number the densities in order of increasing weight, letting $\rho_0=k_1$, $\rho_1=k_2$ and
so on; thus, the density in~\eqref{B1kevQ} is denoted by $\rho_3$, since its weight falls between
those of $\rho_2$ and $\rho_4$.
\end{rem}

The curve f\/lows~\eqref{B1curv} and~\eqref{B2curv} turn out to share the same conservation laws;
for example, $\int  k_1\,\dx$ is conserved by~\eqref{B1curv} because~\eqref{B1kev} implies that
\begin{gather*}
(k_1)_t=D(k_1'+2k_2).
\end{gather*}
Similarly,~\eqref{B2curv} conserves $\int  k_2\,\dx$ because~\eqref{B2kev} implies that
\begin{gather*}
(k_2)_t=D\left(\tfrac23k_1^{(4)}+k_2'''-2k_1k_1''-(k_1')^2-2k_1k_2'+\tfrac49k_1^3+2k_2^2\right).
\end{gather*}
In Section~\ref{Five} we will show that these f\/lows share an inf\/inite sequence of conservation
laws.

\subsection{Lax representation} In this subsection we use geometric considerations to derive Lax
pairs for curvature evolution equations induced by~\eqref{B1curv} and~\eqref{B2curv}.

In~\cite{CIM} we found that the components of the solution $\gamma(x,t)$ of Pinkall's f\/low
satisf\/ied the scalar Lax pair for the KdV equation.
In the same spirit, we seek a~system of the form
\begin{gather}
\label{laxrep}
\L y=0,
\qquad
y_t=\M y,
\end{gather}
satisf\/ied by each component of $\gam$, where $\L$ and $\M$ are dif\/ferential operators in $x$
with coef\/f\/icients involving $k_1$, $k_2$.
Using~\eqref{gamthird}, we see that every component of $\gam$ satisf\/ies the scalar ODE
$y'''=(k_1y)'+k_2y$, and so we will let
\begin{gather*}
\L:=D^3-D k_1-k_2
\end{gather*}
and seek operators $\M_1$ for~\eqref{B1curv} and $\M_2$ for~\eqref{B2curv}.

In the case of~\eqref{B1curv}, the components of $\gam$ also satisfy $y_t=y''-\tfrac23k_1y$, so we
choose
\begin{gather*}
\M_1:=D^2-\tfrac23k_1.
\end{gather*}
One can then verify that~\eqref{B1kev} implies that
\begin{gather}
\label{B1lax}
\L_t=[\M_1,\L].
\end{gather}

In the case of~\eqref{B2curv}, the components of $\gam$ satisfy $y_t=k_1y''+k_2y'+r_0y$, with $r_0$
as given by~\eqref{B2rzero}.
So, we might set $\M_2=k_1D^2+k_2D+r_0$.
However,~\eqref{laxrep} would also be satisf\/ied if we modify $\M_2$ by adding $\calN\L$, where
$\calN$ is an arbitrary dif\/ferential operator.
In fact, the system~\eqref{B2kev} actually implies that $\L_t=[\M_2,\L]$ for
\begin{gather*}
\M_2:=\big(k_1D^2+k_2D+r_0\big)-3D\L.
\end{gather*}
Writing these systems in Lax form~\eqref{laxrep} enables us to interpolate a~spectral parameter
into the linear equations satisf\/ied by the components.
Thus, consider solutions of the compatible system
\begin{gather}
\label{scalarsystem}
\L y=\lambda y,
\qquad
y_t=\M_j y,
\end{gather}
where $j=1$ or $j=2$.
Of course, the components of the evolving curve satisfy~\eqref{scalarsystem} only when $\lambda=0$.
When $\lambda\not=0$, we can construct solutions of the curve f\/low using solutions
of~\eqref{scalarsystem}:
\begin{prop} Let $k_1$, $k_2$ satisfy the evolution equation~\eqref{B1kev} for $j=1$ or~\eqref{B2kev}
for $j=2$.
For fixed $\lambda\in\R$, let $y_1$, $y_2$, $y_3$ be linearly independent solutions
of~\eqref{scalarsystem}, with Wronskian $W$.
Then $W$ is constant in $x$ and $t$, and $\gam=W^{-1/3}\transpose{(y_1,y_2,y_3)}$ is
arclength-parametrized at each time $t$, with centroaffine invariants $k_1$ and $\tilde
k_2=k_2+\lambda$.
Furthermore, $\gam$ satisfies the evolution equation
\begin{gather*}
\gam_t=
\begin{cases}
 \gam''-\tfrac23k_1\gam,\quad&
j=1,
\\
 k_1\gam''+(\tilde k_2-4\lambda)\gam'+r_0\gam,\quad&
j=2.
\end{cases}
\end{gather*}
\end{prop}
\begin{proof}
If we let $\vy=\transpose{(y_1,y_2,y_3)}$ and form the matrix
$F=(\vy,\vy',\vy'')$, then $F$ satisf\/ies dif\/ferential equations of the form
\begin{gather*}
F^{-1}F_x=
\begin{pmatrix}
0&0&k_2+k_1'+\lambda
\\
1&0&k_1
\\
0&1&0
\end{pmatrix}
,
\qquad
F^{-1}F_t=N_j,
\end{gather*}
where both right-hand side matrices have trace zero.
For example, when $j=1$ one can directly calculate, by dif\/ferentiating $y_t=\M_1y$, that
\begin{gather*}
N_1=
\begin{pmatrix}
-\tfrac23k_1&k_2+\tfrac13k_1'+\lambda&k_2'+\tfrac13k_1''
\vspace{1mm}\\
0&\tfrac13k_1&k_2+\tfrac23k_1'+\lambda
\vspace{1mm}\\
1&0&\tfrac13k_1
\end{pmatrix}.
\end{gather*}
Thus, the Wronskian $W$ is constant in $x$ and $t$.

Because $\gam'''=(k_1\gam)'+(k_2+\lambda)\gam$, the centroaf\/f\/ine invariants of $\gam$ are $k_1$
and $\tilde k_2$.
It is straightforward to compute $\gam_t$ in the $j=1$ case, using $y_t=\M_1y$.
In the $j=2$ case, we compute
\begin{gather*}
y_t=\M_2\,y=k_1y''+(\tilde k_2-\lambda)y'+r_0y-3D\L y=k_1y''+(\tilde k_2-4\lambda)y'+r_0y.\tag*{\qed}
\end{gather*}
\renewcommand{\qed}{}
\end{proof}

\subsection{Connection with Boussinesq equations}
\label{connect}
In~\cite{CQ02} Chou and Qu note that, under the centroaf\/f\/ine curve f\/low~\eqref{B1curv}, the
curvatu\-res~$k_1$,~$k_2$ satisfy a~two-component system of evolution equations that is equivalent to
the Boussinesq equation.
This suggests that the other integrable f\/low~\eqref{B2curv} under discussion may be related to
the Boussinesq hierarchy.

Dickson et al.~\cite{DGU99} write the (f\/irst) Boussinesq equation as a~system
\begin{gather}
\label{Bsq1}
(q_0)_t+\tfrac{1}{6}q_1'''+\tfrac{2}{3}q_1q_1'=0,
\qquad
(q_1)_t-2q_0'=0.
\end{gather}
They embed this in a~hierarchy of integrable equations, each of which is written in Lax form as
\begin{gather}
\label{bsqlax}
L_t=[P_m,L],
\qquad
L:=D^3+q_1D+\tfrac{1}{2}q_1'+q_0,
\end{gather}
where $P_m$ is a~dif\/ferential operator of order $m\not\equiv0\mod3$, with coef\/f\/icients
depending on~$q_0$,~$q_1$ and their $x$-derivatives.
Note that $P_m$ must be chosen so that $[P_m,L]$ has order one.
For example, while $P_1=D$ yields the trivial evolution $(q_1)_t=q_1'$, $(q_0)_t=q_0'$, setting
$P_2=D^2+\tfrac{2}{3}q_1$ gives the Boussinesq equation~\eqref{Bsq1}.

Given the resemblance between~\eqref{B1lax} and~\eqref{bsqlax}, it is tempting to f\/ind
substitutions to connect the Boussinesq equation with~\eqref{B1kev}.
In fact, we can make $L$ and $\L$ coincide by setting
\begin{gather}
\label{ktoqsub}
k_1=-q_1,
\qquad
k_2=\tfrac12q_1'-q_0.
\end{gather}
With this substitution, $\M_1$ coincides with $P_2$, so it follows that~\eqref{B1kev}
and~\eqref{Bsq1} are equivalent.

In~\cite{DGU99} it is shown how the coef\/f\/icients of the operators $P_m$ can be obtained solving
a~recursive system of dif\/ferential equations, and thus these depend on a~number of constants of
integration.
For example, the expression for~$P_4$ is
\begin{gather*}
P_4
= \left[f_1D^2+(g_1-\tfrac{1}{2}f_{1}')D+(\tfrac{1}{6}f_1''-g_1'+\tfrac{2}{3}q_1f_1)\right]+
\\
\hphantom{P_4=}{}
+ \left[f_0D^2+(g_0-\tfrac{1}{2}f_0')D+(\tfrac{1}{6}f_0''-g_0'+\tfrac{2}{3}q_1f_0)\right]L+k_{4,0}+k_{4,1}L,
\end{gather*}
where $f_0=0$, $g_0=1$, $  f_1=\tfrac{1}{3}q_1+c_1$, $  g_1=\tfrac{1}{3}q_0+d_1$, and
$k_{4,0}$, $k_{4,1}$, $c_1$, $d_1$ are arbitrary constants.
For convenience, we will set all these arbitrary constant to zero, so that
\begin{gather*}
P_4=D^4+\tfrac{4}{3}q_1D^2+\tfrac{4}{3}(q_1'+q_0)D+\tfrac{5}{9}q_1''+\tfrac{2}{3}q_0'+\tfrac{2}{9}q_1^2.
\end{gather*}
Again, if we use the substitutions~\eqref{ktoqsub}, we f\/ind that the operator $\M_2$ coincides
with $-3P_4$.
Thus,~\eqref{B2kev} is equivalent to the second nontrivial f\/low in the Boussinesq hierarchy,
provided we also rescale time by $t\to-3t$.

\section{Hierarchies}
\label{Five}

In~\cite{OLie} the Boussinesq hierarchy is discussed as an example of a~bi-Hamiltonian system, in
which two sequences of commuting f\/lows (and conservation laws) are generated by applying
recursion operators.
Thus, given the equivalences established in Section~\ref{connect}, it is not surprising that the
Poisson operators def\/ined in Section~\ref{intsection} can be combined to give a~recursion
operator that generates a~double hierarchy of commuting evolution equations for $k_1$, $k_2$.
In fact, we will show that our recursion operator is equivalent to the Boussinesq recursion
operator as given in Example~7.28 of~\cite{OLie}.
The new information we add is that each of these evolution equations is induced by
a~centroaf\/f\/ine geometric evolution equation for curves, which is itself Hamiltonian relative to
the pre-symplectic structure def\/ined in Section~\ref{sympsec} (see Theorem~\ref{glorias} below).

\subsection{Recursion operators}
\label{curse}
We def\/ine a~sequence of evolution equations for $k_1$, $k_2$
\begin{gather}
\label{kflows}
\dfrac{\di}{\di t_j}
\begin{pmatrix}
k_1
\\
k_2
\end{pmatrix}
=F_j[k_1,k_2],
\end{gather}
via the recursion
\begin{gather}
\label{Fcurs}
F_{j+2}=\P\Q^{-1}F_j,
\end{gather}
with initial data given by
\begin{gather*}
F_0=
\begin{pmatrix}
k_1'
\\
k_2'
\end{pmatrix}
,
\qquad
F_1=
\begin{pmatrix}
k_1''+2k_2'
\\
\tfrac23(k_1k_1'-k_1''')-k_2''
\end{pmatrix}
.
\end{gather*}
(Note that $F_1$ is the right-hand side of~\eqref{B1kev}, while for $j=0$~\eqref{kflows} gives
a~simple transport equation for $k_1$, $k_2$, corresponding to f\/low in the direction of the tangent
vector $\gam'$.)

In order to assert that the $F_j$ def\/ined by~\eqref{Fcurs} are local functions of $k_1$, $k_2$ and
their derivatives~-- i.e., in calculating each $F_j$, the operator $D^{-1}$ is only applied to exact
$x$-derivatives of local functions~-- we cite well-known results on the Boussinesq hierarchy.
For example, the version of the f\/irst Boussinesq equation used by Olver~\cite{OLie} is
\begin{gather}
\label{OBou1}
u_\tau=v',
\qquad
v_\tau=\tfrac13u'''+\tfrac83u u',
\end{gather}
where $\tau$ is the time variable.
If one considers linear transformations on the variables, it is necessary to use some imaginary
coef\/f\/icients to make our version~\eqref{B1kev} of the f\/irst Boussinesq equation for $k_1$, $k_2$
equivalent to~\eqref{OBou1}:
\begin{gather}
\label{kutrans}
x=x,
\qquad
\tau=\ri t,
\qquad
k_1=-2u,
\qquad
k_2=u'-\ri v.
\end{gather}
\begin{prop} Under the above change of variables, the recursion operator $\P\Q^{-1}$ is equivalent
to the Boussinesq recursion operator in~{\rm \cite{OLie}}.
\end{prop} \begin{proof} The transformation between $k_1$, $k_2$ and $u$, $v$ can be written as
\begin{gather*}
\begin{pmatrix}
k_1
\\
k_2
\end{pmatrix}
=\G
\begin{pmatrix}
u
\\
v
\end{pmatrix}
,
\qquad
\G:=
\begin{pmatrix}
-2&0
\\
D&-\ri
\end{pmatrix}
.
\end{gather*}
Thus, if $\di/\di t\,\transpose{(u,v)}=F[u,v]$ is an evolution equation for $u$, $v$, the right-hand
side of the corresponding evolution for $k_1$, $k_2$ is $\G\circ F$.
Thus, our recursion operator $\P\Q^{-1}$ for f\/lows on the~$k_1$,~$k_2$ variables corresponds to
a~recursion operator
\begin{gather}
\label{conjR}
\G^{-1}\P\Q^{-1}\G
\end{gather}
on  f\/lows in the $u$, $v$ variables.
In fact, when one calculates~\eqref{conjR} and substitutes for $k_1$, $k_2$ in terms of $u$, $v$
using~\eqref{kutrans}, the result is exactly $-\ri$ times the Boussinesq recursion operator given
in~\cite{OLie}.
\end{proof}

Since in~\cite{SW} (see Section~5.4 in that paper) it is proven that the Boussinesq recursion
operator from~\cite{OLie} always produces local f\/lows when applied to the `seed' evolution
equations (i.e., the tangent f\/low and f\/irst Boussinesq), it follows that the same is true for
our recursion operator.
\begin{rem} Once one checks that the evolution equations~\eqref{kflows} for $j=0$ and $j=1$
commute, it is automatic from the bi-Hamiltonian structure that all evolution equations in the
sequence~\eqref{kflows} commute in pairs (see, e.g., Theorem 7.24 in~\cite{OLie}).
\end{rem}

It is easy to check that the `seeds' $F_0$, $F_1$ for the recursion are related to the initial
conserved densities by
\begin{gather}
\label{FFvals}
F_0=\P\E\rho_0=\Q\E\rho_2,
\qquad
F_1=\P\E\rho_1=\Q\E\rho_3.
\end{gather}
(The second set of equations was derived in Section~\ref{inty}.) While $\P\Q^{-1}$ is the recursion
operator for commuting f\/lows, it is evident from~\eqref{FFvals} that $\Q^{-1}\P$ should be the
recursion operator for conservation law {\em characteristics} (i.e., the result of applying the
Euler operator $\E$ to a~density).
In fact, we may def\/ine an inf\/inite sequence of conserved densities by
\begin{gather}
\label{rhocurse}
\E\rho_{j+2}=\Q^{-1}\P\E\rho_j,
\qquad
j\ge0.
\end{gather}
The f\/irst few densities calculated using this recursion appear in Table~\ref{rtable}.

\begin{table}[t]  \renewcommand{\arraystretch}{1.30}\caption{}\label{rtable}
\smallskip

{\centering \begin{tabular}{|l|l|}
\hline
Non-stretching vector f\/ield & Conserved density
\\
\hline
$Z_0=\gam'$ & $\rho_0=k_1$
\\
\hline
$Z_1=\gam''-\tfrac23k_1\gam$ & $\rho_1=k_2$
\\[2pt]
\hline
$Z_2=k_1\gam''+k_2\gam'+\ldots$ & $\rho_2=k_1k_2$
\\[2pt]
\hline
$Z_3=(k_1'+2k_2)\gam''+\left(\tfrac13k_1^2-\tfrac23k_1''-k_2'\right)\gam'+\ldots$ &
$\rho_3=\tfrac13(k_1')^2+k_1'k_2+\tfrac19k_1^3+k_2^2$
\\[3pt]
\hline
$Z_4=(-k_1'''-2k_2''+2k_1k_1'+4k_1k_2)\gam''+\Big(\tfrac23k_1^{(4)}+k_2'''$ &
$\rho_4=\tfrac13(k_1'')^2+k_1''(k_2'-k_1^2)-k_1(k_1')^2$
\\
$-2k_1k_1''-(k_1')^2-2k_1k_2'+\tfrac49k_1^3+2k_2^2\Big)\gam'+\ldots$ &
$+(k_2')^2-k_1^2k_2'+\tfrac19k_1^4+2k_1k_2^2$
\\[3pt]
\hline
\end{tabular}

}

\medskip

\noindent
{\small (The coef\/f\/icient of $\gam$ in some vector f\/ields is omitted for reasons of space,
but can be determined from the non-stretching condition.)
\begin{itemize}\itemsep=0pt
\item The $\gam'$ and $\gam''$ coef\/f\/icients of $Z_j$ match
the components of $\E\rho_{j}$.
\item Densities satisfy the recursion relation $\E\rho_{j+2}=\Q^{-1}\P\E\rho_j$
\item $\gam_t=Z_j$ induces curvature evolution $\dbinom{k_1}{k_2}_t=\P\E\rho_{j}=\Q\E\rho_{j+2}$.
\item For $j\ge2$, $Z_j$ is a~Hamiltonian vector f\/ield for $-\int\rho_{j-2}\,\dx$.
\end{itemize}}
\end{table}

We now use these densities to def\/ine a~sequence of f\/lows for centroaf\/f\/ine curves, and
relate each of them to a~curvature evolution equation in the sequence~\eqref{kflows}.
Namely, if $f$ is any local function of $k_1$, $k_2$ and their derivatives, we def\/ine
\begin{gather}
\label{Xsuperdef}
X^f:=(\E f)_1\gam'+(\E f)_2\gam''+r_0\gam,
\end{gather}
where the subscripts indicate the components given by~\eqref{Eulerdef} and $r_0$ is determined by
the non-stretching condition.
Then for the sequence of densities def\/ined recursively by~\eqref{rhocurse} we def\/ine the vector
f\/ields
\begin{gather}
\label{xjdef}
Z_j:=X^{\rho_j}
\end{gather}
and the corresponding sequence of curve f\/lows
\begin{gather}
\label{Hamev}
\gamma_t=Z_j.
\end{gather}
\begin{prop}\sloppy  For each $j\ge0$ the curve flow~\eqref{Hamev} induces the curvature evolution $\frac{\di}{\di
t}\transpose{(k_1,k_2)}=F_j$.
\end{prop} \begin{proof} From~\eqref{FFvals} and the recursion relations, it follows by induction
that
\begin{gather*}
F_j=\P\E\rho_j,
\qquad
j\ge0.
\end{gather*}
Then the result follows immediately from~\eqref{kgenflow}.
\end{proof}

\subsection{Hamiltonian structure at the curve level}
\label{Psect}
We now consider the question of how the Hamiltonian operator $\P$ is related to the Hamiltonian
structure def\/ined at the curve level in Section~\ref{sympsec}.
Recall from Section~\ref{hamex} that $X$ is a~Hamiltonian vector f\/ield associated to the
functional $H$ if
\begin{gather*}
dH[Y]=\omega_\gamma(Y,X)
\end{gather*}
for any non-stretching vector f\/ield $Y$.
\begin{theorem}
\label{glorias}
Let $H(\gam)=\displaystyle\oint_\gamma\rho\,\dx$ and assume that
\begin{gather}
\label{rhatrel}
\E\rhat=\mathcal Q^{-1}\mathcal P\E\rho,
\end{gather}
i.e., $\rhat$ is next after $\rho$ in the sequence of densities generated by the recursion operator
$\Q^{-1}\P$.
Then $-X^{\rhat}$ $($as defined by~\eqref{Xsuperdef}$)$ is Hamiltonian for~$H$.
\end{theorem}
\begin{proof}
Based on the def\/inition~\eqref{curvesymplectic} of $\omega$, we need to show that $\rd H[Y]$ equals
\[
\oint |Y, \gamma', -X^{\widehat\rho}|\rd x = \oint |X^{\widehat\rho}, \gamma', Y|\rd x \qquad \forall\, Y \in T_{\gamma} \widehat M.
\]
If $X=a\gam+b\gam'+c\gam''$ and $Y=\tilde{a}\gam+\tilde{b}\gam'+\tilde{c}\gam''$, then
from~\eqref{XYcomp},
\begin{gather*}
\oint_\gamma|X,\gam',Y|\,\dx=\oint_\gamma(a\tilde{c}-\tilde{a}c)\,\dx.
\end{gather*}
However, using~\eqref{nostretch} to eliminate $a$ and $\tilde{a}$, we obtain
\begin{gather*}
\oint_\gamma|X,\gam',Y|\,\rd x=
\oint_\gamma\left(-b'\tilde{c}+c\,\tilde{b}'+\tfrac{1}{3}(c\,\tilde{c}''-c''\tilde{c})\right)\rd x=
-\oint_\gamma(b'\tilde{c}+c'\tilde{b})\,\rd x,
\end{gather*}
where the last equation follows by integration by parts.
Thus,
\begin{gather*}
\oint_\gamma\big|X^{\rhat},\gam',Y\big|\dx=-\oint_\gamma
\begin{pmatrix}
\bt
\\
\ct
\end{pmatrix}
\cdot\mathcal\Q\,\E\rhat\,\dx=-\oint_\gamma
\begin{pmatrix}
\bt
\\
\ct
\end{pmatrix}
\cdot\P\E\rho\,\dx,
\end{gather*}
using~\eqref{rhatrel} in the last step.
Then, because $\P$ is skew-adjoint,
\begin{gather*}
\oint_\gamma\big|X^{\rhat},\gam',Y\big|\,\dx=\oint_\gamma\E\rho\cdot\mathcal P
\begin{pmatrix}
\bt
\\
\ct
\end{pmatrix}
\dx.
\end{gather*}
On the other hand, using the properties of the Euler operator we have
\begin{gather*}
dH[Y]=\oint_\gamma(\E\rho)_1\delta_Y k_1+(\E\rho)_2\delta_Y k_2\,\dx,
\end{gather*}
where $\delta_Y$ denotes the f\/irst variation in the direction of $Y$.
Now using~\eqref{kgenflow} we have
\begin{gather*}
dH[Y]=\oint_\gamma\E\rho\cdot
\begin{pmatrix}
\delta_Y k_1
\\
\delta_Y k_2
\end{pmatrix}
=\oint_\gamma\E\rho\cdot\P
\begin{pmatrix}
\bt
\\
\ct
\end{pmatrix}
\dx.
\end{gather*}
This concludes the proof.
\end{proof}

The following corollaries are immediate consequences of the theorem.
\begin{cor} Define the Poisson bracket
\begin{gather*}
\{H,G\}=\oint \E h\cdot\mathcal P\E g\,\dx,
\end{gather*}
where $G(k_1, k_2) = \oint g\, \rd x$ and $H(k_1, k_2) = \oint h \,\rd x$, and $g$, $h$ are functions of periodic $k_1$, $k_2$ and their derivatives.
Then $dH[X^g]=\{H,G\}$.
\end{cor}

\begin{cor} The vector fields $Z_j$ defined by~\eqref{xjdef} are Hamiltonian for $j\ge2$.
\end{cor}

\begin{cor} A closed curve $\gam$ is critical for the functional
$H(\gam)= \oint_{\gamma}\rho_j\,\rd x$ with respect to non-stretching variations if and
only if $\gam$ is stationary for a~constant-coefficient linear combination
$Z_{j+2}+c_0Z_0+c_1Z_1$.
\end{cor}

\begin{proof} By Theorem~\ref{glorias}, $-Z_{j+2}$ is Hamiltonian for $H$.
Thus, a~curve $\gam$ is $H$-critical if and only if $\omega(Y,Z_{j+2})=0$ for all $Y\in
T_\gamma\Mhat$.
This condition is satisf\/ied if and only if $Z_{j+2}$ is in the kernel of $\omega_\gamma$, i.e.,
along $\gam$ it is equal to a~constant-coef\/f\/icient linear combination of vectors $Z_0$ and~$Z_1$.
Equivalently, $Z_{j+2}+c_0Z_0+c_1Z_1=0$ along $\gam$ for some constants $c_0$, $c_1$, expressing the
property that $\gam$ is stationary for a~linear combination of these vector f\/ields.
\end{proof}

\subsection{Projective properties}
\label{projsec}
As stated in Remark~\ref{projremark}, a~centroaf\/f\/ine curve $\gam$ projects to give a~conic in
$\RP^2$ if and only if the Wilczynski invariants satisfy $p_0-\tfrac12p_1'=0$.
(The corresponding condition in terms of $k_1$, $k_2$ is $k_2+\tfrac12k_1'=0$.) In this subsection we
will investigate f\/lows in the hierarchy having the property that, if $\gam$ projects to a~conic
at time zero, then it continues to have a~conical projection at subsequent times.
We will show later that the equation of the conic in homogeneous coordinates is f\/ixed in time.
We will also discuss f\/lows that preserve a~parametrization that is proportional to projective
arclength; in that case, the corresponding condition in terms of curvatures is that
$k_2+\tfrac12k_1'$ is a~nonzero constant along the curve.

These investigations are much easier if, instead of $k_2$, we use an invariant that vanishes
precisely when the condition we are investigating holds.
Accordingly, we f\/ix a~constant $C$, and def\/ine an alternative pair of invariants
\begin{gather}
\label{altvals}
u=k_1,
\qquad
v=k_2+\tfrac12k_1'-C.
\end{gather}
Thus, the curve is a~conic if $v=0$ when $C=0$, and the curve has a~constant-speed parametrization
(relative to projective arclength) if $v=0$ when $C\ne0$.

We will convert the evolution equations in the hierarchy (at the level of the invariants) to these
variables.
Suppose that a~curve f\/low causes the invariants to evolve by
\begin{gather*}
\begin{pmatrix}
k_1
\\
k_2
\end{pmatrix}
_t=F[k_1,k_2],
\end{gather*}
where $F$ is a~vector-valued function of $k_1$, $k_2$ and their derivatives.
Then the corresponding evolution equation for the alternative invariants is
\begin{gather*}
\begin{pmatrix}
u
\\
v
\end{pmatrix}
_t=\G\circ F\big[u,v-\tfrac12u'+C\big],
\qquad
\G:=
\begin{pmatrix}
1&0
\\
\tfrac12D&1
\end{pmatrix}
.
\end{gather*}

Similarly, if $\calR$ is the recursion operator generating the hierarchy of evolution equations for~$k_1$,~$k_2$, then the recursion operator for the corresponding f\/lows on $u$, $v$ dif\/fers by a~gauge
transformation:
\begin{gather*}
\Rt=\G\calR\G^{-1},
\qquad
\text{where}
\qquad
\G^{-1}=
\begin{pmatrix}
1&0
\\
-\tfrac12D&1
\end{pmatrix}
.
\end{gather*}
(It is understood that, in $\calR$ on the right-hand side, $k_1$, $k_2$ are substituted for in terms
of~$u$,~$v$.) Specif\/ically, using $\calR=\P\Q^{-1}$ as def\/ined by~\eqref{defofP},~\eqref{defofQ},
we compute
\begin{gather*}
\Rt=\Rt_0+\Ft_0D^{-1}
\begin{pmatrix}
0&1
\end{pmatrix}
+\Ft_1D^{-1}
\begin{pmatrix}
1&0
\end{pmatrix}
,
\end{gather*}
where
\begin{gather*}
\Rt_0=
\begin{pmatrix}
3(v+C)&2\big(u-D^2\big)
\\
\calN&3(v+C)
\end{pmatrix}
,
\qquad
\Ft_0=
\begin{pmatrix}
u'
\\
v'
\end{pmatrix}
,
\qquad
\Ft_1=
\begin{pmatrix}
2v'
\\
\tfrac23uu'-\tfrac16u'''
\end{pmatrix}
,
\end{gather*}
and $\calN$ is the scalar dif\/ferential operator $\tfrac16D^4-\tfrac56u
D^2-\tfrac54u'D+\tfrac23u^2-\tfrac34u''$.
One can check that the vectors $\Ft_0$, $\Ft_1$ are the time derivatives of $u$, $v$, corresponding to the
`seeds' $Z_0$ and $Z_1$ for the hierarchy of curve f\/lows.

By applying $\Rt$ to $\Ft_0$, $\Ft_1$, one can generate the right-hand sides of the evolution
equations in the hierarchy in terms of~$u$ and~$v$.
Letting $\Ft_j$ denote these vectors, we compute (for example)~that
\begin{gather*}\Ft_2  = \Rt
\Ft_0 =
\begin{pmatrix}
-2v'''+4(uv)'+3C u'
\vspace{1mm}\\
\tfrac16 u^{(5)}-uu'''-2u'u''+\tfrac43 u^2 u' +(4 v +3C) v'
\end{pmatrix}
,
\\
\Ft_3  =\Rt \Ft_1 =
\begin{pmatrix}
\tfrac13 u^{(5)} +\tfrac53(u^2 u' - u u''')-\tfrac{25}6 u'u'' + (10 v +6C)v'
\vspace{1mm}\\
\tfrac13 v^{(5)}-\big(\tfrac56 v +\tfrac12 C\big)u''' +\tfrac53 ((u^2 v)'-u v'''-u'' v') - \tfrac52
u'v'' +2Cu u'
\end{pmatrix}
.
\end{gather*} Here, when applying $D^{-1}$ to dif\/ferential polynomials in $u$, $v$, the constant of
integration is taken to be zero.

Notice in particular that if $v\equiv0$, then the bottom component of $\Ft_3-3C\Ft_1$ vanishes.
Thus, the f\/low $Z_3-3C Z_1$ preserves the condition that $v$ is identically zero.
In fact, we can calculate two inf\/inite sequences of evolution equations for $u$, $v$ that preserve
this condition; the right-hand sides of these are
\begin{gather}
\label{nicepat}
G_k=\begin{cases}{\displaystyle\sum_{j=0}^k}\binom{k}{j}(-3C)^j\Ft_{2(k-j)},&\text{$k$ even},
\vspace{1mm}\\
{\displaystyle\sum_{j=0}^k}\binom{k}{j}(-3C)^j\Ft_{2(k-j)+1},\quad &\text{$k$ odd}.
\end{cases}
\end{gather}
While it is routine to verify that any individual curvature evolution equation in these sequences
preserves $v\equiv0$, it is easier to observe that the members of these sequences satisfy the
recursion relation
$G_{k+2}=(\Rt-3C)^2G_k$.
Then the fact that they all preserve $v\equiv0$ is a~consequence of the following:
\begin{prop}
\label{bootstrap}
If a~curvature evolution $(u_t,v_t)^{\rm T}=G_k[u,v]$ in this sequence preserves $v\equiv0$, then so does
the evolution $(u_t,v_t)^{\rm T}=G_{k+2}[u,v]$.
\end{prop} \begin{proof} We assume that $G_k=\transpose{(D\ell_1,D\ell_2)}$ for local functions
$\ell_1$, $\ell_2$ of $u$, $v$ and their derivatives.
(This form for~$G_k$ is necessary if we are able to apply operator $\Rt$ to it and produce local
functions.)

Within the ring of polynomials in $u$, $v$ and their derivatives, let $\cV$ denote the ideal generated
by~$v$, $v'$, $v''$, etc.
By hypothesis, $D\ell_2\in\cV$, and the same is true for $\ell_2$.

We compute
\begin{gather}
\label{oneR}
(\Rt-3C)G_k=\ell_1\Ft_1+\ell_2\Ft_0+
\begin{pmatrix}
3v D\ell_1+2\big(u-D^2\big)D\ell_2
\\
\calN D\ell_1+3v D\ell_2
\end{pmatrix}
.
\end{gather}
Thus, the bottom component of $(\Rt-3C)^2G_k$ is given by
\begin{gather*}
\big(\calN+\Ft_{12}D^{-1}\big)\big(\ell_1\Ft_{11}+\ell_2\Ft_{01}+3v D\ell_1+2(u-D^2)D\ell_2\big)
\\
\qquad
{} +\big(3v+\Ft_{02}D^{-1}\big)\big(\ell_1\Ft_{12}+\ell_2\Ft_{02}+\calN D\ell_1+3v D\ell_2\big).
\end{gather*}
(Here, $\Ft_{j1}$ and $\Ft_{j2}$ denote the top and bottom entries in the vector $\Ft_j$.) The second factor in the top line is the top entry of $(\Rt-3C)G_k$.
This polynomial lies in~$\cV$, and the same is true if we apply $\calN$ or $D^{-1}$ to it.
On the other hand, because $\Ft_{02}=v'\in\cV$, the coef\/f\/icient in front on the second line
also vanishes when $v\equiv0$.
\end{proof}

In the special case when $C=0$, we see that the following curve f\/lows (as def\/ined
by~\eqref{xjdef}) preserve conicity:
\begin{gather}
\label{Clist}
Z_0,
\quad
Z_3,
\quad
Z_4,
\quad
Z_7,
\quad
Z_8,
\quad
Z_{11},
\quad
Z_{12},
\quad
\ldots.
\end{gather}
However, when $C\ne0$, some care needs to be taken in matching the evolution equations for~$u$,~$v$
that preserve $v\equiv0$ with the corresponding linear combinations of the curve f\/lows~$Z_j$.

\looseness=-1
In the proof of Proposition~\ref{bootstrap}, we used the fact that if an exact derivative $D\ell$ lies in $\cV$, then by
choosing the constant of integration equal to zero, the antiderivative $\ell$ also lies in $\cV$.
However, if we express $D\ell$ in terms of $k_1$, $k_2$ instead of $u$, $v$, and then take an
antiderivative, a~particular constant of integration must be chosen in order to belong in $\cV$.
Thus, when we compute the $k$th evolution equation for $u$, $v$ by using the recursion operator $\Rt$
(which involves applying $D^{-1}$), then convert this to an evolution equation for~$k_1$,~$k_2$, and
f\/inally try to match it with a~curve f\/low in the hierarchy~\eqref{Hamev}, we get a~linear
combination of $Z_k$ with lower-order f\/lows of the same parity.
For example, if we substitute~\eqref{altvals} into~$\Ft_2$, and then apply the operator~$\G^{-1}$,
we~get
\begin{gather*}
\G^{-1} \Ft_2\big[ k_1, k_2 \!+\! \tfrac12 k_1' \!\!-\! C\big]   =
\!\!
\begin{pmatrix}
-k_1''''\!-\!2k_2'''\!+\!2k_1 k_1'' \!+\! 4 (k_1 k_2)' \!+\! 2(k_1')^2 \!-\! C k_1'
\\[2pt]
\tfrac23 k_1''''' \!+\! k_2''''\!-\!2 k_1 k_1''' \!-\! 4 k_1' k_1'' \!+\! 4 k_2 k_2' \!-\!2(k_1 k_2')' \!+\! \tfrac43
(k_1)^2 k_1' \!-\! C k_2'
\end{pmatrix}
\\
\hphantom{\G^{-1} \Ft_2\big[ k_1, k_2 \!+\! \tfrac12 k_1' \!\!-\! C\big]}{}
 =  F_2 - C F_0.
\end{gather*}
Thus, $\Ft_2$ is induced by the curve f\/low $Z_2-C Z_0$; similarly, $\Ft_3$ is
induced by $Z_3-2C Z_1$, and so~on.

Similarly, when we apply the recursion operator $(\Rt-3C)^2$ to generate higher-order f\/lows that
preserve $v\equiv0$, these constants of integration accumulate and change the relatively nice
pattern of the coef\/f\/icients exhibited by~\eqref{nicepat}.
Here is what we get when we compute the f\/irst few curve evolutions corresponding to the
evolutions $G_k$:
\begin{center}
\begin{tabular}{|l|l|}
\hline
$u$, $v$ evolution & centroaf\/f\/ine curve f\/low \phantom{\Big|}
\\
\hline $G_0$ & $Z_0$
\\
$G_1$ & $Z_3-5C Z_1$
\\
$G_2$ & $Z_4-7C Z_2+14C^2Z_0$
\\
$G_3$ & $Z_7-11C Z_5+44C^2Z_3-\frac{220}{3}C^3Z_1$
\\[3pt]
$G_4$ & $Z_8-13C Z_6+65C^2Z_4-\frac{455}{3}C^3Z_2+\frac{455}{3}C^4Z_0$
\\[3pt]
\hline
\end{tabular}
\end{center}

\subsection[Conical evolutions and the Kaup-Kuperschmidt hierarchy]{Conical evolutions and the Kaup--Kuperschmidt hierarchy}

\looseness=-1
In this section we examine special properties of the conicity-preserving f\/lows~\eqref{Clist}
(hence, from now on we are assuming $C=0$).
These properties will enable us to connect our hierarchy of centroaf\/f\/ine curve f\/lows in
$\R^3$ with the Kaup--Kuperschmidt hierarchy and with curve f\/lows in centroaf\/f\/ine $\R^2$.
We begin with the observation that, when restricted to conical curves, the coef\/f\/icient of
$\gam''$ vanishes for as many of the vector f\/ields in~\eqref{Clist} as one cares to check.
In other words, this coef\/f\/icient belongs to $\cV$, the ideal within the ring of dif\/ferential
polynomials in $k_1$, $k_2$ generated by $k_2+\tfrac12k_1'$ and its derivatives.
In fact, this is true in general, as shown in the following:

\begin{prop}
\label{nor2}
If $j\equiv0$ or $j\equiv-1$ modulo $4$, then the bottom component of $\E\rho_j$ belongs in~$\cV$;
hence, the $\gam''$ coefficient of $Z_j$ vanishes on conical curves.
\end{prop}

\begin{proof} The statement can be verif\/ied directly for $j=0$ and $j=3$.
For higher values, we use the recursion relation between the characteristics, which implies that
$\E\rho_{j+4}=(\Q^{-1}\P)^2\E\rho_j$.
From Proposition~\ref{bootstrap} we know that the bottom component of $\Ft_j=\G\P\E\rho_j$ lies in $\cV$.
By inserting powers of $\G$ and $\G^{-1}$ into the recursion relation, we get
\begin{gather*}
\E\rho_{j+4}=\Q^{-1}\P\Q^{-1}\G^{-1}\G\P\E\rho_j=\Q^{-1}\calR\G^{-1}\Ft_j.
\end{gather*}
As in the proof of Proposition~\ref{bootstrap}, we can assume that $\Ft_j=(D\ell_1,D\ell_2)^{\rm T}$ where $\ell_2\in\cV$.
Taking $C=0$ in equation~\eqref{oneR}, we see that the top entry of
$\calR\G^{-1}\Ft_j=\G^{-1}\Rt\Ft_j$ is
\begin{gather}
\label{gallot}
\ell_1\Ft_{11}+\ell_2\Ft_{01}+3v D\ell_1+2\big(u-D^2\big)D\ell_2,
\end{gather}
which clearly is in $\cV$.
(Note that $v=k_2+\tfrac12k_1'$ here.) Noting the form of $\Q$, we see that the bottom entry of
$E\rho_{j+4}$ is $D^{-1}$ applied to~\eqref{gallot}, so it must also belong to $\cV$.
\end{proof}

Next, we make a~connection with curve f\/lows in centroaf\/f\/ine $\R^2$ to show that, for the
f\/lows in~\eqref{Clist}, the cone that the curve lies on is preserved by the time evolution.
\begin{prop}
If $\gam(x,t)$ evolves by any of the vector fields in~\eqref{Clist}, and $\gam(x,0)$
lies on a~cone through the origin in $\R^3$, then $\gam(x,t)$ lies on the same cone at later times.
\end{prop}
\def\fV{\mathsf{V}}
\begin{proof}
Using the action of $\SL(3)$ we can, without loss of
generality, assume that the equation of the cone is $y_1y_3-(y_2)^2=0$.
We f\/ix a~map $\fV$ from $\R^2$ onto this cone:
\begin{gather*}
\fV
\begin{pmatrix}
x_1
\\
x_2
\end{pmatrix}
=2^{-1/3}
\begin{pmatrix}
x_1^2
\\
x_1x_2
\\
x_2^2
\end{pmatrix}
.
\end{gather*}
Of course, when we projectivize on each end this gives the {\em Veronese embedding} of $\RP^1$ as
a~quadric in $\RP^2$.
The scale factor of $2^{-1/3}$ is chosen so that if $X(x)$ is a~parametrized curve in $\R^2$
satisfying the centroaf\/f\/ine normalization $|X,X'|=1$, then $\Gamma=\fV\circ X$ satisf\/ies the
normalization $|\Gamma,\Gamma',\Gamma''|=1$.
Moreover, if $p(x)$ is the curvature of $X$, then the invariants of $\Gamma$ are $k_1=-4p$ and
$k_2=2p'$.
Finally, if $X$ evolves by the non-stretching f\/low
\begin{gather}
\label{nonstr2}
X_t=r X'-\tfrac12r'X,
\end{gather}
then $\Gamma(x,t)=\fV\circ X(x,t)$ satisf\/ies
$
\Gamma_t=r\Gamma'-r'\Gamma$.

By Proposition~\ref{nor2} all the f\/lows in~\eqref{Clist}, when restricted to conical curves, are of this form,
for some choice of dif\/ferential polynomial $r$ in $k_1$.
For any initial data $\gam(x,0)$, we can def\/ine a~curve $X(x,0)$ in $\R^2$ such that
$\gam(x,0)=\fV\circ X(x,0)$ and make $X(x,t)$ evolve by~\eqref{nonstr2}.
Because $\Gamma(x,t)=\fV\circ X(x,t)$ satisf\/ies the same initial value problem, then
$\gam(x,t)=\Gamma(x,t)$ at all times, and $\gam(x,t)$ lies on the cone def\/ined by
$y_1y_3-(y_2)^2=0$ at all times.
\end{proof}

In~\cite{CQ02} Chou and Qu discovered a~non-stretching f\/low for curves in centroaf\/f\/ine
$\R^3$ which preserves the conicity condition $k_2+\tfrac12k_1'=0$ and which causes the curvature
$k_1$ to evolve by the Kaup--Kuperschmidt equation:
\begin{gather*}
u_t=u'''''-5u u'''-\tfrac{25}2u'u''+5u^2u'
\end{gather*}
(see Case 3 in Section~3 of their paper, taking $\lambda=0$).
In fact, up to a~multiplicative factor of~$1/3$, Chou and Qu's f\/low is the same as the
restriction of $Z_3$ to conical curves.

Not only does f\/low $Z_3$ give a~geometric realization of the Kaup--Kuperschmidt equation, but the
entire sequence~\eqref{Clist} of f\/lows realizes the Kaup--Kuperschmidt hierarchy, when restricted
to conical curves.
To see this, note that the {\em square} of the recursion operator $\Rt$ relates the evolution of
$k_1=u$ under $Z_j$ to its evolution under $Z_{j+4}$.
(Here, we use the notation of Section~\ref{projsec} but with $C=0$ and $v=0$ because of conicity.)
The resulting recursion operator is
\begin{gather*}
-\tfrac13D^6+2u D^4+6u'D^3+\left(\tfrac{49}{6}u''\!-\!3u^2\right)D^2+\left(\tfrac{35}{6}u'''\!-\!10u u''\right)D
+\tfrac{13}{6}u''''-\tfrac{41}{6}u u''
\\
\qquad
{} -\tfrac{23}{4}(u')^2+\tfrac{4}{3}u^3+u'D^{-1}\circ(\tfrac{1}{3}u^2\!-\!\tfrac{1}{6}u'')
+\tfrac13\left(u'''''\!-\!5u u'''\!-\!\tfrac{25}2u'u''\!+\!5u^2u'\right)D^{-1}.
\end{gather*}
This agrees with the known recursion operator for symmetries of the Kaup--Kuperschmidt hierar\-chy.
(See, e.g., Example~2.20 in~\cite{JPlist}, where the operator dif\/fers by changing~$u$ to~$-u$ and
rescaling time by a~factor of~$1/3$.) Using this, one can check that the curvature f\/lows induced
by~\eqref{Clist} for conical curves coincide with the commuting f\/lows of the Kaup--Kuperschmidt
hierarchy.
\begin{rem} The curve f\/low discovered by Chou and Qu is in fact also def\/ined for
centroaf\/f\/ine curves parametrized proportional to projective arclength (i.e., those for which
$k_2+\tfrac12k_1'$ is a~constant), and nevertheless still induces Kaup--Kuperschmidt evolution for
$k_1$.
Recently, Musso~\cite{musso} has extended this to a~hierarchy of f\/lows for arclength-parametrized
curves in $\RP^2$ which induces the Kaup--Kuperschmidt hierarchy for curvature evolution.
We suspect that these f\/lows coincide with the restrictions of the f\/lows studied in
Section~\ref{projsec} (for $C\ne0$) to the centroaf\/f\/ine lifts of such curves in $\RP^2$.
\end{rem}
\begin{rem} Schwartz and Tabachnikov~\cite{ST} showed that certain maps def\/ined on the space of
convex polygons preserve the subset of polygons that are inscribed (or circumscribed) on a~conic:
that is, if the vertices of the polygon (or those of its projective dual) lie on a~conic, then the
same is true for its image under the map.
The building blocks for these maps are elementary maps $T_r$ that associate to a~given polygon
another polygon obtained from the intersections of diagonals joining each vertex to the vertex
located $r$ positions to the left or right of it.
In fact, the maps preserving conicity are particular combinations of $T_r$ for certain values of $r$.

The map corresponding to $r=2$ is called the {\em pentagram map} and it is known to be
a~discretization (in both time and space) of the Boussinesq equation~\cite{OST}.
It is natural to wonder if the maps in~\cite{ST} are somehow associated to f\/lows in the
Boussinesq hierarchy.
We are currently investigating this.
\end{rem}

\subsection*{Acknowledgements}

The authors would like to acknowledge the careful and detailed work of the anonymous referees in
helping improve this paper.
The authors also gratefully acknowledge support from the National Science Foundation: A.~Calini
through grants DMS-0608587 and DMS-1109017, and as a~current NSF employee; T.~Ivey through grant
DMS-0608587; and G.~Mar\'i~Bef\/fa through grant DMS-0804541.
T.~Ivey also acknowledges support from the College of Charleston Mathematics Department.
G.~Mar\'\i~Bef\/fa also acknowledges the support of the Simons Foundation through their Fellows
program.

\pdfbookmark[1]{References}{ref}
\LastPageEnding

\end{document}